\documentclass[twocolumn,showpacs,preprintnumbers,amsmath,amssymb,
superscriptaddress,nofootinbib]{revtex4-1}
\usepackage{graphicx}

\newcommand{\eps}{\varepsilon}
\newcommand{\tens}[1]{{\boldsymbol{#1}}}

\newcommand{\pa}{{\tens{\partial}}}
\newcommand{\lied}{\pounds}

\newcommand{\n}[1]{\label{#1}}
\newcommand{\be}{\begin{equation}}
\newcommand{\ee}{\end{equation}}
\newcommand{\ba}{\begin{eqnarray}}
\newcommand{\ea}{\end{eqnarray}}

\newcommand{\beq}{\begin{equation}}
\newcommand{\eeq}{\end{equation}}
\newcommand{\beqa}{\begin{eqnarray}}
\newcommand{\eeqa}{\end{eqnarray}}

\newcommand{\hook}{\raisebox{-0.35ex}{\makebox[0.6em][r]
{\scriptsize $-$}}\hspace{-0.15em}\raisebox{0.25ex}{\makebox[0.4em][l]{\tiny
 $|$}}}
\newcommand{\eq}[1]{(\ref{#1})}

\begin{document}

\title{Generalized Killing--Yano equations in $D=5$ gauged supergravity}

\author{David Kubiz\v n\'ak}

\email{D.Kubiznak@damtp.cam.ac.uk}

\affiliation{DAMTP, University of Cambridge, Wilberforce Road, Cambridge CB3 0WA, UK}

\author{Hari K. Kunduri}

\email{H.K.Kunduri@damtp.cam.ac.uk}

\affiliation{DAMTP, University of Cambridge, Wilberforce Road, Cambridge CB3 0WA, UK}

\author{Yukinori Yasui}

\email{yasui@sci.osaka-cu.ac.jp}

\affiliation{Department of Mathematics and Physics, Graduate School of Science, Osaka City University, 3-3-138 Sugimoto, Sumiyoshi, Osaka 558-8585, JAPAN}

\date{June 16, 2010}  

\begin{abstract}
We propose a generalization of the (conformal) Killing--Yano equations relevant to $D=5$ minimal gauged supergravity. The generalization stems from the fact that
the dual of the Maxwell flux, the 3-form $\tens{*F}$, couples naturally to particles in the background as a `torsion'.
Killing--Yano tensors in the presence of torsion preserve most of the properties of the standard Killing--Yano tensors---exploited recently for the higher-dimensional rotating black holes of vacuum gravity with cosmological constant. In particular,
the generalized closed conformal Killing--Yano 2-form gives rise to the tower of generalized closed conformal Killing--Yano tensors of increasing rank which in turn generate the tower of Killing tensors. An example of a generalized Killing--Yano tensor is found for the Chong-Cveti\'c-L\"u-Pope black hole spacetime [hepth/0506029]. Such a tensor stands behind the separability of the Hamilton--Jacobi, Klein--Gordon, and Dirac equations in this background.
\end{abstract}

\pacs{04.65.+e, 04.50.-h, 02.40.-k, 04.50.Gh, 04.20.Jb \hfill  DAMTP-2009-38, OCU-PHYS 314}

\maketitle

\section{Introduction}
In recent years there has been considerable progress in the study of  properties of the most general known stationary higher-dimensional vacuum (including a cosmological constant) black hole spacetimes with spherical horizon topology~\cite{ChenEtal:2006cqg}. This progress builds on the
discovery that, as in four dimensions, these spacetimes admit a closed conformal Killing--Yano 2-form \cite{FrolovKubiznak:2007, KubiznakFrolov:2007}. From this tensor one can generate the towers of explicit and hidden symmetries \cite{KrtousEtal:2007jhep} which underline the
complete integrability of geodesic motion \cite{PageEtal:2007, KrtousEtal:2007prd}, separability of the Hamilton--Jacobi \cite{FrolovEtal:2007}, Klein--Gordon \cite{FrolovEtal:2007}, Dirac \cite{OotaYasui:2008, Wu:2008, Wu:2008b, AhmedovAliev:2009}, and gravitational \cite{KunduriEtal:2006, OotaYasui:2009} perturbations, and integrability of stationary string equations \cite{KubiznakFrolov:2008} in these backgrounds.   The most general metric element admitting a closed conformal Killing--Yano 2-form was constructed in \cite{HouriEtal:2007, KrtousEtal:2008, HouriEtal:2008b, HouriEtal:2009a}. When this 2-form is non-degenerate and the vacuum Einstein equations are imposed, the unique solution is the metric describing the Kerr-NUT-(A)dS spacetime \cite{ChenEtal:2006cqg}. Hence the study of Killing--Yano tensors has been important in establishing a uniqueness result.

Although such progress is impressive it is limited to vacuum spacetimes and therefore excludes important examples of black holes with non-trivial gauge fields, such as those of various supergravity theories which arise in low-energy limits of string theory compactifications. The vacuum black holes of~\cite{ChenEtal:2006cqg} have charged generalizations in various dimensions (see \cite{EmparanReall:2008} and references therein). An important question is whether Killing-Yano tensors exist for these charged solutions, and further if one can prove uniqueness results in analogy with the vacuum case. Although these spacetimes have been shown to admit integrable geodesic motion (see \cite{Chow:2008, DavisEtal:2005} and references therein) the existence of a deeper structure, such as Killing--Yano tensors, remains to be shown. Note that the Kerr-Newman black hole, which may be viewed as a solution of $D=4$ supergravity, has a Killing--Yano tensor (i.e its square is the associated Killing tensor) that satisfies the same equation as the Kerr solution. This appears, however, to be a special property; for example, in five dimensions, one can easily find a `square root' of the Killing tensors of charged black holes, but this does not solve the `vacuum' Killing--Yano equation.

The first step to fill this gap was achieved recently by Wu \cite{Wu:2009}. Wu demonstrated
that the Dirac equation in the background of the most general known charged rotating spherical black hole of $D=5$ minimal supergravity \cite{CveticYoum:1996} (it has been proved this is the unique such solution with $R \times U(1)^2$ isometry~\cite{TomizawaEtal:2009}) can be separated provided that an extra counterterm is added to this equation, and that this separability can be justified by the existence of a symmetry operator related to a `generalized' Killing--Yano 3-form. It can be expected that the same remains true also in the presence of a cosmological constant \cite{Wu:2009}.

The aim of this paper is to provide a geometric understanding of the existence of generalized Killing--Yano tensors in spacetimes with gauge fields.  Specifically, we propose a generalization of the (conformal) Killing--Yano equations for the theory of $D=5$ minimal gauged supergravity. Our definition is motivated by the study of non-generic symmetries in the motion of classical spinning particles in the presence of a torsion field and the fact that the Hodge dual of the Maxwell field strength $\tens{*F}$ of this theory couples naturally to particles as an extra torsion term. In particular, we demonstrate that when the 3-form $\tens{*F}$ is identified with the torsion, the most general known spherical black hole solution of a minimal gauged supergravity, the Chong--Cveti\'c--L\"u--Pope black hole \cite{ChongEtal:2005b}, admits a generalized closed conformal Killing--Yano 2-form. Due to various algebraic identities valid in this spacetime, the equation for this 2-form reduces to the one found by Wu in the absence of the cosmological constant. Moreover, viewing the generalized Killing--Yano equation from the point of view of torsion naturally explains why an additional counterterm has to be added to the Dirac equation so that this is separable, and gives an explicit expression for the symmetry operator allowing this separability.

The paper is organized as follows: In Section 2 we study a modification of Killing--Yano equations in the presence of a totally antisymmetric torsion field.  In Section 3, we review  the theory of $D=5$ minimal gauged supergravity
and propose a generalization of Killing--Yano equations for this theory. In Section 4,
we demonstrate that the Chong--Cveti\'c--L\"u--Pope black hole admits a generalized closed conformal Killing--Yano 2-form.
Section 5 is devoted to discussion. The theory of classical spinning particles relevant to our work is summarized in Appendix A.

\section{Killing--Yano tensors in the presence of torsion}
Let $\tens{T}$ be a 3-form on a $D$-dimensional Riemannian manifold $(M,\tens{g})$ and
$\{\tens{e}_a \}$ an orthonormal frame, $\tens{g}(\tens{e}_a, \tens{e}_b)=\delta_{ab}$.
Let us define a connection $\tens{\nabla}^T$ by (we sum over $a=1\dots D$)
\be\label{j1}
\nabla^T_X \tens{Y}=\nabla_X \tens{Y}+\frac{1}{2}\, \tens{T}(\tens{X}, \tens{Y},
\tens{e}_a) \tens{e}_a\,,
\ee
where $\tens{X}, \tens{Y}$ are vector fields and $\tens{\nabla}$ is a Levi-Civita connection. This connection
satisfies a metricity condition, $\nabla^T_X \tens{g}=0$, and has the same geodesics
as $\tens{\nabla}$, $\nabla^T_{\dot{\gamma}} \tens{\dot{{\gamma}}}=\nabla_{\dot{\gamma}} \tens{\dot{\gamma}}=0$ for a geodesic
$\gamma$. The connection 1-form $(\tens{\omega}^T)^a_{~b}$ is introduced by
\be\label{j2}
\nabla^T_{e_b} \tens{e}_a= (\tens{\omega}^T)^c_{~a}(\tens{e}_b) \,\tens{e}_c\,.
\ee
Comparing \eqref{j1} and \eqref{j2} we have
\beq\label{j3}
\tens{\omega}^T_{ab}=\tens{\omega}_{ab}-\frac{1}{2}T_{abc}\tens{e}^c\,,
\eeq
where $\tens{\omega}_{ab}$ is the Levi-Civita connection 1-form,
\beq\label{j4}
\tens{\omega}_{ab}=-\tens{\omega}_{ba}\,,\quad \tens{de}^a+\tens{\omega}^a_{~b} \wedge \tens{e}^b=0\,.
\eeq
The 1-form $\tens{\omega}^T_{ab}$ satisfies
\beq\label{j4}
\tens{\omega}^T_{ab}=-\tens{\omega}^T_{ba}\,,\quad \tens{de}^a+
(\tens{\omega}^T)^a_{~b} \wedge \tens{e}^b = \tens{T}^a\,,
\eeq
where $\tens{T}_a(\tens{X},\tens{Y})=\tens{T}(\tens{e}_a, \tens{X}, \tens{Y})$\,.

For a $p$-form $\tens{\Psi}$ we calculate a covariant derivative as\footnote{We use notations of \cite{FrolovKubiznak:2008}. In particular, the `hook' operator $\hook$ corresponds to the inner derivative.}
\be\label{j5}
\nabla^T_X \tens{\Psi}=\nabla_X \tens{\Psi}-\frac{1}{2}
\bigl(\tens{X}\hook \tens{e}_b\hook \tens{T}\bigr)\wedge (\tens{e}_b\hook \tens{\Psi})\,.
\ee
Then, we have
\ba
\tens{d}^T \tens{\Psi}\!&=&\!\tens{e}^a \wedge \nabla^T_{e_a} \tens{\Psi}\nonumber\\
\!&=&\!\tens{d\Psi}-(\tens{e}_a\hook \tens{T}) \wedge (\tens{e}_a\hook\tens{\Psi})\,,
\quad\ \ \label{j6}\\
\tens{\delta}^T \tens{\Psi}\!&=&\!-\tens{e}_a\hook\nabla^T_{e_a}\tens{\Psi}\nonumber\\
\!&=&\!\tens{\delta \Psi}-\frac{1}{2}\bigl(\tens{e}_a\hook \tens{e}_b\hook \tens{T}\bigr) \wedge \bigl(\tens{e}_a\hook \tens{e}_b\hook \tens{\Psi}\bigr)\,.\quad\ \ \label{j7}
\ea
For $\tens{\Psi}=\tens{T}$ one has $\tens{\delta}^T \tens{T}=\tens{\delta T}$.
In the case of 5-dimensional manifold and
a 3-form $\tens{\Psi}$, \eqref{j6} is written as
\beq\label{use}
\tens{d}^T \tens{\Psi}
=\tens{d \Psi}-(\tens{*T}) \wedge (\tens{*\Psi})\,.
\eeq

We define a {\em generalized conformal Killing-Yano} (GCKY) tensor $\tens{k}$ to be a $p$-form satisfying
for any vector field $\tens{X}$
\beq\label{CKY}
\nabla^T_X \tens{k}-\frac{1}{p+1}\tens{X}\hook \tens{d}^T \tens{k}+\frac{1}{D-p+1}\tens{X}^\flat \wedge \tens{\delta}^T \tens{k}=0\,.
\eeq
In analogy with Killing--Yano tensors defined with respect to the Levi-Civita connection, we call a GCKY tensor $\tens{f}$ obeying
$\tens{\delta}^T \tens{f}=0$ a {\em generalized Killing--Yano} (GKY) tensor, and a GCKY
$\tens{h}$ obeying
$\tens{d}^T \tens{h}=0$ a {\em generalized closed conformal Killing--Yano} (GCCKY) tensor.
Basic properties of GCKY tensors are gathered in the following lemma:\\
{\bf Lemma.}
(1) A GCKY 1-form is equal to a conformal Killing 1-form.
(2) The Hodge star $\tens{*}$ maps GCKY $p$-forms into GCKY $(D-p)$-forms. In particular, the Hodge star of a GCCKY $p$-form is a GKY $(D-p)$-form and vice versa.
(3) When $\tens{h}_1$ and $\tens{h}_2$ is a GCCKY $p$-form and $q$-form, respectively, the wedge product $\tens{h}_1 \wedge \tens{h}_2$ is a GCCKY $(p+q)$-form.
(4) A symmetric tensor $\tens{K}$ constructed from a GKY $p$-form $\tens{f}$ as
\beq
\tens{K}(\tens{X},\tens{Y})=\tens{g}(\tens{X}\hook\tens{f}, \tens{Y}\hook\tens{f})
\eeq
is a Killing tensor.\\
{\em Proof}. The property (1) is easy to see, (2) follows from the commutability
$\tens{*\nabla}^T=\tens{\nabla}^T \tens{*}$, the proof of (3) proceeds in the same way as in the case of (standard) Killing--Yano tensors \cite{KrtousEtal:2007jhep}, see also \cite{FrolovKubiznak:2008}. Let us prove (4).
For any vector fields $\tens{X}, \tens{Y}, \tens{Z}$ we have
\ba
(\nabla^T_X \tens{K})(\tens{Y},\tens{Z})\!&=&\!\!\tens{g}(\tens{Y}\hook \nabla^T_X \tens{f},\tens{Z}\hook\tens{f})\nonumber\\
\!&+&\!\!\tens{g}(\tens{Y}\hook \tens{f}, \tens{Z}\hook \nabla^T_X \tens{f})\nonumber\\
\!&=&\!\!\frac{1}{p+1}\,\tens{g}(\tens{Y}\hook \tens{X}\hook \tens{d}^T \tens{f},
\tens{Z}\hook \tens{f})\nonumber\\
\!&+&\!\!\frac{1}{p+1}\,\tens{g}(\tens{Z}\hook\tens{X}\hook \tens{d}^T \tens{f},\tens{Y}\hook\tens{f})\,,\qquad\nonumber
\ea
and
\beqa
(\nabla^T_X \tens{K})(\tens{Y}, \tens{Z})\!&=&\!(\nabla_X \tens{K})(\tens{Y},\tens{Z})\nonumber\\
\!&-&\!\frac{1}{2} \tens{T}(\tens{X},\tens{Y}, \tens{e}_a)\tens{K}(\tens{e}_a, \tens{Z})\nonumber\\
\!&-&\!\frac{1}{2}\tens{T}(\tens{X},\tens{Z}, \tens{e}_a)\tens{K}(\tens{Y}, \tens{e}_a)\,.\quad
\nonumber
\eeqa
Thus we have
\be
\nabla_X \tens{K}(\tens{Y}\!,\!\tens{Z})\!+\!\nabla_Z \tens{K}(\tens{X},\!\tens{Y})\!+\!\nabla_Y \tens{K}(\tens{Z},\!\tens{X})\!=\!0\,.\ \Box
\ee
As in the case of closed conformal Killing--Yano tensors, properties (2)--(4) are very useful for generating other GCCKY and Killing tensors.
In particular, a GCCKY 2-form gives rise to the tower of
GCCKY tensors and the tower of Killing tensors, exactly in the same way as without torsion \cite{KrtousEtal:2007jhep, FrolovKubiznak:2008}.

\section{Minimal gauged supergravity and generalized Killing--Yano equations}
The bosonic sector of $D=5$ minimal gauged supergravity is governed by the Lagrangian
\be\label{L}
\pounds=\tens{*}(R+\Lambda)-\frac{1}{2}\tens{F}\wedge \tens{*F}\!+
\frac{1}{3\sqrt{3}}\,\tens{F} \wedge \tens{F}\wedge \tens{A}\,.
\ee
This yields the following set of Maxwell and Einstein equations:
\ba
\tens{dF}=0\,,\quad \tens{d* F}-\frac{1}{\sqrt{3}}\,\tens{F}\wedge\tens{F}\!&=&\!0\,,\quad
\label{F}\label{Maxwell}\\
R_{ab}-\frac{1}{2}\Bigl(F_{ac}F_b^{\ c}-\frac{1}{6}\,g_{ab}F^2\Bigr)+\frac{1}{3}\Lambda g_{ab}\!&=&\!0\,.
\label{Einstein}
\ea

In our further analysis we adopt an assumption that the Maxwell flux 3-form
$\tens{*F}$ plays the role of torsion. In particular, we associate
\be\label{TF}
\tens{T}=\frac{1}{\sqrt{3}}\, \tens{*F}\,.
\ee
Having done so, we first notice that the {\em torsion is ``harmonic''} with respect to the torsion covariant derivative,
\be
\tens{\delta}^T\tens{T}=0\,, \quad \tens{d}^T\tens{T}=0\,.
\ee
The first equality follows from the fact that $\tens{\delta}^T\tens{T}=\tens{\delta T}$ and the exactness of $\tens{F}$, the second from the equality \eqref{use} and the Maxwell equations \eqref{Maxwell}.
Note that the above construction will work even if the prefactor multiplying the Chern-Simons term in the action is an arbitrary constant. However, in this work we will be concerned only with supergravity, for which explicit solutions are known.\footnote{%
Recently, a uniqueness theorem for the most general known, charged rotating black hole solution with spherical horizon in the minimal ungauged $D=5$ supergravity was demonstrated \cite{TomizawaEtal:2009}. An important ingredient in the proof is the precise value of the Chern--Simons coupling. For black hole solutions with other values of
this coefficient see \cite{KleihausEtal:2008, AlievCiftci:2009}.}

We propose that the generalized conformal Killing--Yano equations for the $D=5$ minimal gauged supergravity are given by \eqref{CKY} with the torsion identified in \eqref{TF}.
In particular, let us focus on a GCCKY 2-form $\tens{h}$. For such a 2-form we get the following equation:
\be\label{PCKY1}
\nabla_X^{T}\tens{h}=\tens{X}^{\flat}\wedge \tens{\xi}\,,\quad
\tens{\xi}=-{1\over 4}\,\tens{\delta}^{T}\tens{h}\, ,
\ee
which can be rewritten in components as
\be\label{PCKY2}
\nabla_c h_{ab}=2g_{c[a}\xi_{b]}-\frac{1}{\sqrt{3}}\,(*F)_{cd[a}h^d_{\ \,b]}\,.
\ee
Of course, the 2-form $\tens{h}$ is $\tens{d}^T$-closed; $\tens{d}^T\tens{h}=0$. The requirement that it is also closed, $\tens{d}\tens{h}=0$,
imposes the following algebraic condition on the flux $\tens{*F}$:
\be\label{con1}
(*F)_{d[ab}h^d_{\ \,c]}=0\,\ \Leftrightarrow \ F^c_{\ [a}h_{b]c}=0\,.
\ee
When this is satisfied, there exists (locally) a potential 1-form $\tens{b}$ such
that $\tens{h}=\tens{db}$.
Similarly, when a condition
\be\label{con2}
(*F)_{abc} h^{bc}=0\ \Leftrightarrow \ \tens{F}\wedge \tens{h}=0
\ee
is satisfied we have $\tens{\delta}^T \tens{h}=\tens{\delta} \tens{h}$.
In any case, lemma in Section 2 implies that $\tens{h}$ gives rise to a Killing tensor $\tens{K}$,
\be\label{KT}
K_{ab}=(*h)_{acd}(*h)_b^{\ cd}=h_{ac}h_b^{\ c}-\frac{1}{2}g_{ab}h^2\,,
\ee
and at least one Killing 1-form $\tens{\eta}$, $\tens{\eta}=\tens{*h}^{\wedge 2}$.

The question about the existence of other Killing vectors is an interesting one.
Let us focus on a 1-form $\tens{\xi}$ which in the vacuum case turns out to be a
primary Killing 1-form. Repeating the construction in \cite{KrtousEtal:2008} we find
$\lied_\xi \tens{h}=0$. Moreover, by contracting the integrability conditions for Eq. \eqref{PCKY1} we can show that $\nabla_{a}\xi^{a}\propto R^T_{[ab]}h^{ba}=0$ due to the fact that torsion $\tens{T}$ is ``harmonic'', and 
\be\label{bohuzel}
\nabla_{(a}\xi_{b)}=\frac{1}{D-2}\Bigl\{R^T_{c (a}h_{b)}{}^{\!c}
-R^T_{c (a b) d} h^{cd}\Bigr\}\,.
\ee
When condition \eqref{con2} is satisfied, the second term on the r.h.s. vanishes.
The vanishing of the first term imposes an algebraic condition on $\tens{F}$ and $\tens{h}$,
$F_a{}^d F_{d(b}h_{c)}{}^a=0$. 

In the following section we give an explicit example of black hole spacetime
admitting a non-degenerate GCCKY 2-form for which this condition as well as
both conditions \eqref{con1} and \eqref{con2} are satisfied.

\section{An example}
The most general known black hole solution of $D=5$ minimal gauged supergravity, constructed by
Chong, Cveti\v{c}, L\"u, and Pope \cite{ChongEtal:2005b}, can be written in the following orthonormal form, cf. \cite{LuEtal:2008b}:
\ba
\tens{g}\!&=&\!\sum_{\mu=x,y}\bigl(\tens{\omega}^{\mu}\tens{\omega}^{\mu}+
\tens{\tilde \omega}^{\mu}\tens{\tilde\omega}^{\mu}\bigr)
+\tens{\omega}^{\eps}\tens{\omega}^{\eps}\,,\label{can_odd}\\
\tens{A}\!&=&\!\sqrt{3}(\tens{A}_q+\tens{A}_p)\,.\label{A}
\ea
Here,
\ba\label{omega}
\tens{\omega}^{x} \!\!&=&\!\sqrt{\frac{x-y}{4X}}\,\tens{d}x\,,\quad
\tens{\tilde \omega}^{x}=\frac{\sqrt{X}(\tens{d}t+y\tens{d}\phi)}{\sqrt{x(y-x)}}\,,\nonumber\\
\tens{\omega}^{y} \!\!&=&\!\!\sqrt{\frac{y-x}{4Y}}\,\tens{d}y\,,\quad
\tens{\tilde \omega}^{y}=\frac{\sqrt{Y}(\tens{d}t+x\tens{d}\phi)}{\sqrt{y(x-y)}}\,,\\
\tens{\omega}^{\eps}\!\!\!&=&\!\!\!\frac{1}{\sqrt{\!-\!xy}}\bigl[\mu \tens{d}t\!+\!\mu(x\!+\!y)\tens{d}\phi\!+\!xy\tens{d}\psi\!-\!y\tens{A}_q
\!-\!x\tens{A}_p\bigr],\nonumber\\
\tens{A}_q\!\!&=&\!\!\frac{q}{x-y}(\tens{d}t+y\tens{d}\phi)\,,\quad\tens{A}_p=\frac{-p}{x-y}(\tens{d}t+x\tens{d}\phi)\,,\nonumber
\ea
and metric functions take the form
\ba
X\!&=&\!(\mu+q)^2+Ax+Cx^2+\frac{1}{12}\Lambda x^3\,,\nonumber\\
Y\!&=&\! (\mu+p)^2+By+Cy^2+\frac{1}{12}\Lambda y^3\,\label{XY}.
\ea The solution has $R \times U(1)^2$ isometry group generated by the Killing fields $\pa_t, \pa_\psi, \pa_\phi$.
We have used a `symmetric' gauge; the Maxwell field strength $\tens{F}=\tens{dA}$ depends on $Q=q-p$. Moreover, one can use the translations in $x$ and $y$ to eliminate $C$, leaving total number of four free parameters characterizing the solution. These are related to two independent angular momenta, mass, and electric charge.

The spacetime possesses a non-degenerate GCCKY 2-form $\tens{h}$ obeying \eqref{PCKY1}. Tensors $\tens{F}$ and $\tens{h}$ are such that both conditions \eqref{con1} and \eqref{con2}
are satisfied. Hence, $\tens{\delta}^T \tens{h}=\tens{\delta h}$ and $\tens{h}$ can be generated from a potential $\tens{b}$.\footnote{%
In fact, it turns out that conditions \eqref{con1} and \eqref{con2}
(with $\tens{h}=\tens{db}$) are so strong that they
completely determine $\tens{h}$ in this spacetime.}
The potential and the GCCKY 2-form read
\ba
\tens{b}\!&=&\!-\frac{1}{2}\Bigl[(x+y)\tens{d}t+xy\tens{d}\phi\Bigr]\,,\\
\tens{h}\!&=&\!\tens{db}=\sqrt{-x}\, \tens{\tilde \omega}^{x}
\wedge \!\tens{\omega}^{x}+\sqrt{-y}\, \tens{\tilde \omega}^{y}
\wedge \!\tens{\omega}^{y}\,.\quad\label{h_odd}
\ea
The second equation means that basis $\{\tens{\omega}\}$ is a Darboux basis for $\tens{h}$ and coordinates $x, y$ are the eigenvalues of $\tens{h}^2$.

The 2-form $\tens{h}$ has very similar properties to the closed conformal Killing--Yano tensor of the 5D Kerr-NUT-AdS spacetime \cite{FrolovKubiznak:2007, KubiznakFrolov:2007}. Namely, it gives rise to the
Killing tensor, \eqref{KT},
\ba\label{K}
\tens{K}\!&=&\!y(\tens{\omega}^{x}\tens{\omega}^{x}+
\tens{\tilde \omega}^{x}\tens{\tilde\omega}^{x})
+x(\tens{\omega}^{y}\tens{\omega}^{y}+
\tens{\tilde \omega}^{y}\tens{\tilde\omega}^{y})\quad \nonumber\\
\!&+&\!(x+y)\tens{\omega}^{\eps}\tens{\omega}^{\eps}\,,
\ea
and implies the existence of two isometries
\be
\tens{\xi}=(\pa_{t})^\flat\,,\quad
\tens{*}(\tens{h}^{\wedge 2})=2(\pa_{\psi})^\flat\,.
\ee
However, one finds
\be
K^{ab}\xi_b=(\partial_{\phi})^a+\frac{p-q}{x-y}(\partial_{\psi})^a\,.
\ee
This means that, up to the fact that we do not recover $\pa_\phi$, coordinates $(t, \phi, \psi,x ,y)$ are `canonical coordinates', completely
determined by $\tens{h}$ (see, e.g., \cite{FrolovKubiznak:2008} for more details in the vacuum case).

The Hamilton--Jacobi,  Klein--Gordon, and Dirac equations
in the background of the Chong--Cveti\v{c}--L\"u--Pope black hole were studied in \cite{DavisEtal:2005}. It is straightforward to check that
the Hamilton--Jacobi and Klein--Gordon equations allow separation of variables. This is directly related
to the existence of Killing tensor \eqref{K}.  However, the separability of the Dirac equation was achieved only in the special case in which the two independent angular momenta of the black hole are set equal.
On the other hand, Wu recently demonstrated \cite{Wu:2009} that
the Dirac equation separates in the general rotating (ungauged) black hole background provided that an extra counter term proportional to
$(*F)_{abc}\gamma^{a}\gamma^{b}\gamma^c$ is added to it.
Although this counter term seems apparently `strange', such a modification of the Dirac operator is quite natural from the point of view of torsion which we are discussing here.
Indeed, referring to the Appendix A we find that the proper commuting symmetry operators
in this background are expected to be ($\tens{f}\equiv\tens{*h}$)
\ba
\hat Q\!&=&\!\gamma^a\nabla_a^s-\frac{1}{24\sqrt{3}}\gamma^{a}\gamma^{b}\gamma^{c}(*F)_{abc}\,,\\
\hat Q_{f}\!\!&=&\! \gamma^{a}\gamma^{b}f^c_{\ ab}\nabla_c^s
\!+\!\frac{1}{32}\gamma^{a}\gamma^{b}\gamma^{c}\gamma^{d}
W_{abcd}\,,\qquad\n{Qf1}
\ea
where
\be
W_{abcd}=
(df)_{abcd}-\frac{2}{\sqrt{3}}T^e_{\ [ab}f_{|e|cd]}=(df)_{abcd}\,.
\ee
The second equality is not true for a general $\tens{f}$ but follows from the constraint~\eqref{con2}.
The latter operators correspond to the symmetry operators found by Wu \cite{Wu:2009}.
We also note that the necessary condition \eq{necessary} is satisfied.

\section{Discussion}
Let us summarize our results. In this paper we have studied generalized conformal Killing--Yano equations in $D=5$ minimal gauged supergravity.
The generalization we have proposed  stems from the torsion-like behavior of the dual of the Maxwell 2-form $\tens{*F}$. The identification of $\tens{*F}$ with torsion has many appealing features, in particular, the Maxwell equations imply that the torsion is ``harmonic''.
The GCKY tensors possess many of the properties of the standard conformal Killing--Yano tensors. For example, a GCCKY 2-form generates a Killing tensor and at least one Killing vector.

We have presented an explicit example of a spacetime admitting a
GCCKY 2-form $\tens{h}$. The metric describes
the most general known black hole solution of
$D=5$ minimal gauged supergravity, the Chong--Cveti\'c--L\"u--Pope black hole
\eqref{can_odd}--\eqref{XY}. The tensor $\tens{h}$ in this example is further constrained to obey
conditions \eqref{con1} and \eqref{con2}. It gives rise to the Killing tensor and
two isometries. An interesting question is whether the last isometry can also be exploited from the existence of $\tens{h}$. If this is the case, the GCCKY tensor determines
all the canonical coordinates and one can hope to prove that the Chong--Cveti\'c--L\"u--Pope black hole is the unique solution which possesses this additional structure (similar to the uniqueness of the Kerr-NUT-(A)dS spacetime in the case of standard closed conformal Killing--Yano tensor \cite{HouriEtal:2007, KrtousEtal:2008}).

The main significance of Killing--Yano tensors is that they allow one to construct
symmetry operators allowing the separability of various field equations.
Our GKY tensors proceed in the same line.
The GCCKY 2-form $\tens{h}$ is responsible for the separability of the Hamilton--Jacobi and Klein--Gordon equations. It also uniquely
determines the modified Dirac equation for which the variables can be separated.
The question of separability for higher spins is, similar to the vacuum case, open.

Another issue is a relationship of generalized hidden symmetries and algebraic type of solutions.
Whereas the existence of a standard non-degenerate conformal Killing--Yano 2-form
limits spacetime to the type D \cite{MasonTaghavi:2008}  of higher-dimensional classification
\cite{ColeyEtal:2004a, Coley:2008}, our explicit example of Chong--Cveti\'c--L\"u--Pope black hole (which is type~I) demonstrates that this is no longer true
for the GCKY tensors.
Following the calculation in Appendix C.1 of \cite{Kubiznak:phd}, one can easily show that the (non-zero eigenvalue) eigenvectors of a non-degenerate GCCKY 2-form are null geodesics.
In our example, these are also Weyl aligned. Is this (as in the vacuum case
\cite{HamamotoEtal:2007}) true in general? What is the most general type of spacetime admitting a GCKY 2-form?

The results obtained in this paper raise many interesting questions.
In the case of proposed GCKY tensors these are related to the uniqueness,
separability of equations with higher spin, relationship of algebraic type and hidden
symmetries. A more fundamental question is: is it possible to generalize Killing--Yano equations for other types of supergravity theories? And if so, how many of the key properties of the standard Killing--Yano tensors are preserved by solutions to such equations? A natural starting point would be to consider $U(1)^3$ gauged supergravity, for which explicit solutions are already known, e.g., the supersymmetric solution found in \cite{KunduriEtal:2006b}. 
The satisfactory answer to these questions could bring many new insights into analytical properties of various black hole spacetimes.

\appendix

\section{Spinning particles in the presence of torsion}

It is well known that Killing--Yano tensors correspond to the existence of an enhanced worldline sypersymmetry of the theory of classical spinning particles \cite{GibbonsEtal:1993}. This fact provides a systematic tool for derivation of (generalized) Killing--Yano equations in the presence of various types of fields. For example, it turns out that in the presence of an electromagnetic field the Killing--Yano equations do not get modified; only an additional algebraic condition is imposed on an electromagnetic 2-form  \cite{Tanimoto:1995}. This is no longer true in the presence of torsion \cite{RietdijkvanHolten:1996}, or a torsion-like 3-form in string theory \cite{DeJongheEtal:1996}.
As discussed in the main text, the Maxwell--Chern--Simons 3-form $\tens{*F}$ of the minimal gauged supergravity behaves effectively like a torsion. That is why in this appendix we review the
theory of classical spinning particles in the presence of torsion and study its non-generic supersymmetries connected with generalized Killing--Yano tensors.

The explicit derivation of the equation for the Killing--Yano 2-form in the presence of torsion was performed by Rietdeik and Holten \cite{RietdijkvanHolten:1996}.
Following closely their paper, we shall generalize this equation for an arbitrary
p-form.
As usual, we denote particle's worldline coordinates by $x^\mu$ and describe its spin
by a pseudo Lorentz vector of Grassmann-odd coordinates $\psi^a$. In the presence
of an (antisymmetric) torsion $T_{abc}$ the
Lagrangian reads
\cite{Rietdijk:1992, RietdijkvanHolten:1996}
\ba\label{theory}
L&=&\frac{1}{2}\,g_{\mu\nu}\dot x^\mu \dot x^\nu-\frac{i}{2}\eta_{ab}\dot \psi^a\psi^b
\nonumber\\
&-&\frac{i}{2}\dot x^\mu\psi^a\psi^b(\omega_{\mu ab}+\frac{1}{2}T_{\mu ab})\\
&-&\frac{1}{2}\frac{1}{4!}\psi^a\psi^b\psi^c\psi^d (dT)_{abcd}\,.\nonumber
\ea
This theory possesses a generic supercharge $Q$,
\be
Q=\psi^a e_a^{\ \mu}\Pi_\mu-\frac{i}{6}\psi^a\psi^b\psi^c T_{abc}\,,
\ee
which obeys
\be
\{H,Q\}=0\,, \quad \{Q,Q\}=-2iH\,.
\ee
Here, $H$ is the Hamiltonian,
\ba
H\!&=&\!\frac{1}{2}\Pi_\mu\Pi_\nu g^{\mu\nu}+\frac{1}{2}\frac{1}{4!}\psi^a\psi^b\psi^c\psi^d (dT)_{abcd}\,,\qquad\\
\Pi_\mu\!&=&\!p_\mu+\frac{i}{2}\psi^a\psi^b(\omega_{\mu ab}+\frac{1}{2}T_{\mu ab})\,,
\ea
$e_a^{\ \mu}$ denotes the vielbein,
$p_\mu$ is the momentum conjugate to $x^\mu$ and the Poisson brackets are defined as
\be\label{brackets}
\{F,G\}=\frac{\partial F}{\partial x^\mu}\frac{\partial G}{\partial p_\mu}-
\frac{\partial F}{\partial p_\mu}\frac{\partial G}{\partial x^\mu}+
i(-1)^{a_F}\frac{\partial F}{\partial \psi^a}\frac{\partial G}{\partial \psi_a}\,,
\ee
where $a_F$ is the Grassmann parity of $F$. For practical calculations it is useful to
rewrite these brackets in a covariant form
\ba\label{brackets2}
\{F,G\}\!&=&\!D^T_\mu F\frac{\partial G}{\partial \Pi_\mu}-
\frac{\partial F}{\partial \Pi_\mu}D_\mu^T G\nonumber\\
\!&+&\!(T_{\mu\nu}^{\ \ \kappa}\Pi_\kappa-\frac{i}{2}\psi^a\psi^b R_{ab\mu\nu}^T\bigr)\frac{\partial F}{\partial \Pi_\mu}\frac{\partial G}{\partial \Pi_\nu}\nonumber\\
\!&+&\!i(-1)^{a_F}\frac{\partial F}{\partial \psi^a}\frac{\partial G}{\partial \psi_a}\,,
\ea
where we have used the phase space covariant derivative with torsion
\ba
D^T_\mu F\!&=&\!\partial_\mu F+(\omega_{\mu ab}+\frac{1}{2}T_{\mu ab})\psi^b\frac{\partial F}{\partial \psi_a}\nonumber\\
\!&+&\!(\Gamma_{\mu\nu}^{\ \ \kappa}+\frac{1}{2}T_{\mu\nu}^{\ \ \,\kappa})\Pi_\kappa\frac{\partial F}{\partial \Pi_\nu}\,
\ea
and $R_{ab\mu\nu}^T$ is the Riemann tensor with torsion.

The non-generic supercharge $Q_{f}$ is derived from the equation
\be\label{Q}
\{Q, Q_{f}\}=0\,.
\ee
From the Jacobi identity it follows that such a supercharge is automatically a constant of motion $\{H, Q_f\}=0$. Using explicit expressions for $Q$ and $H$ these equations read
\ba
\{H,Q_f\}\!&=&\!\!\frac{1}{2}\frac{1}{4!}\psi^a\psi^b\psi^c\psi^d\nabla_\mu^T (dT)_{abcd}
\frac{\partial Q_f}{\partial \Pi_\mu}\nonumber\\
&-&\!\!\Pi^\mu D^T_\mu Q_f-\frac{i}{2}\psi^a\psi^b R_{ab\mu\nu}^T\Pi^\mu\frac{\partial Q_f}{\partial \Pi_\nu}
\nonumber\\
&+&\!\frac{i}{12}(dT)_{abcd}\psi^b\psi^c\psi^d\frac{\partial Q_f}{\partial \psi_a}\quad
\label{kam1}\\
&=&0\,,\nonumber\\
\{Q,Q_f\}&=&\!-\frac{i}{6}\psi^a\psi^b\psi^c\nabla_\mu^T T_{abc}
\frac{\partial Q_f}{\partial \Pi_\mu}-\psi^\mu D^T_\mu Q_f\nonumber\\
&+&\!\!\bigl(T_{\mu\nu}^{\ \ \kappa}\Pi_\kappa-\frac{i}{2}\psi^a\psi^b R_{ab\mu\nu}^T\bigr)\psi^\mu\frac{\partial Q_f}{\partial \Pi_\nu} \nonumber\\
&-&\!i\bigl(\Pi^a-\frac{i}{2}T^a_{\ \,bc}\psi^b\psi^c\bigr)\frac{\partial Q_f}{\partial \psi^a}\quad\nonumber\\
&=&0\,.\label{kam2}
\ea
In particular, we use the following ansatz for a non-generic supercharge:
\ba\label{ansatz}
Q_f\!&=&\!\psi^{a_1}\!\dots \psi^{a_{p-1}} f_{a_1\dots a_{p-1}}{}^\mu\Pi_{\mu} \nonumber\\
\!&+&\!\frac{i}{(p+1)!}\psi^{a_1}\!\dots \psi^{a_{p+1}} \tilde c_{a_1\dots a_{p+1}}\,.\quad
\ea
When plugged into Eqs. \eqref{kam1} and \eqref{kam2} we get equations for various components characterized by number of fermionic and momentum coordinates, $\{\#_{\psi},\#_{\Pi}\}$,
which have to vanish separately. Specifically, component $\{p-1,2\}$ of Eq. \eqref{kam1}
implies
\be
\nabla^T_{(\mu} f_{|\alpha_1\dots \alpha_{p-1}|\nu)}=0\,,
\ee
Components $\{p-2,2\}$ and $\{p,1\}$ of \eqref{kam2} imply $f_{\alpha_1\dots \alpha_p}=f_{[\alpha_1\dots \alpha_p]}$, and
\ba
\!\!\!&&\nabla^T_{[\mu}f_{\alpha_1\dots \alpha_{p-1}]\lambda}
-\frac{p-1}{2}T_{[\mu}{}^\sigma{}_{\alpha_1}f_{|\sigma|\alpha_2\dots \alpha_{p-1}]\lambda}
\nonumber\\
&&-T_{[\mu}{}^\sigma{}_{|\lambda|}f_{\alpha_1\dots \alpha_{p-1}]\sigma}
-\frac{1}{p!}\tilde c_{\lambda\mu\alpha_1\dots\alpha_{p-1}}=0\,.\quad\ \ 
\ea
From here we can obtain a simplified expression for $\tens{\tilde c}\,,$
\ba\label{ctilde}
\frac{(-1)^{p+1}}{(p+1)!}\tilde c_{\mu\lambda\alpha_1\dots\alpha_{p-1}}\!&=&\!
-\frac{1}{p+1}\nabla^T_{[\mu}f_{\lambda \alpha_1\dots \alpha_{p-1}]}\quad\nonumber\\
\!\!\!\!\!&+&\!\frac{1}{2}T_{[\mu}}{}^\sigma{}_\lambda
f_{|\sigma|\alpha_1\dots \alpha_{p-1}]\,,\ \ 
\ea
and the equation for the $p$-form $\tens{f}$ corresponding to $f_{\alpha_1\dots\alpha_{p}}$.
In the language of forms the generalized Killing--Yano equation in the presence
 of torsion reads
\be\label{lll}
\nabla^T_X \tens{f}=\frac{1}{p+1}\tens{X\hook d^T f}\,.
\ee
The remaining components of Eqs. \eq{kam1} and \eq{kam2} give the following condition on torsion:
\be\label{necessary}
\tens{d}(\tens{d}^T\!\tens{f})-\frac{p+1}{2}(\tens{e}^a\hook \tens{dT})\wedge (\tens{e}_a\hook \tens{f})=0\,,
\ee 
which is a $p$-form generalization of the condition (21) derived in \cite{DeJongheEtal:1996} (see also (4.16) in \cite{HouriEtal:2010a}).  Only if this condition is satisfied,
the quantity $Q_f$, \eq{ansatz}, where $\tens{f}$ is a generalized Killing--Yano tensor obeying \eq{lll} and $\tens{\tilde c}$ is given by 
\eq{ctilde},  represents a non-generic supercharge of the theory \eq{theory}.

Let us finally address the question why the Dirac equation has to be modified in order to allow the separation of variables. For a transition from classical to quantum theory of
spinning particles see, e.g, \cite{Rietdijk:1992, Cariglia:2004}.
Using the (naive) correspondence relations
\be
\Pi_\mu\to -i\bigl(\nabla_\mu^{s}-\frac{1}{8}\gamma^{ab}T_{\mu ab}\bigr)\,,\quad \psi^\mu\to \frac{1}{\sqrt{2}}\gamma^\mu\,,
\ee
where $\nabla_\mu^{s}=\partial_\mu-\frac{1}{4}\gamma^{ab}\omega_{\mu ab}$ is a spinorial covariant derivative,
we find that the operators corresponding to $Q$ and $Q_f$ are the following
modified Dirac and symmetry operators:
\ba
\hat Q\!&=&\!\gamma^\mu\nabla_\mu^s-\frac{1}{24}T_{abc}\gamma^{a}\gamma^b\gamma^c\,,\n{Dirac}\\
\hat Q_f\!&=&\! \gamma^{a_1}\!\dots\! \gamma^{a_{p-1}}f^\mu{}_{a_1\dots a_{p-1}}\nabla_\mu^s\nonumber\\
&\!+\!&\frac{1-p}{8(p+1)}\gamma^{a_1}\!\dots\!\gamma^{a_{p+1}}T^b_{\ [a_1a_2}
f_{|b| a_3\dots a_{p+1}]}\nonumber\\
&\!+\!&\frac{1}{2(p+1)^2}\gamma^{a_1}\!\dots\!\gamma^{a_{p+1}}(df)_{a_1\dots a_{p+1}}\,.\n{Qf1}
\ea
Up to quantum anomalies, these operators commute (anticommute) for $p$ even (odd).
Although in the case of standard Killing--Yano tensors these anomalies cancel \cite{Cariglia:2004}, there are 
additional anomalies in the case with torsion. These were recently calculated and can be found in \cite{HouriEtal:2010a}.

\vspace*{3ex}
\section*{Acknowledgments}
\vspace*{-1ex}
We are grateful to G.W. Gibbons and C. Warnick for discussions.
D.K. acknowledges the Herchel Smith Postdoctoral Research Fellowship at the University of Cambridge. H.K.K. is an STFC Postdoctoral Research Fellow.


\end{document}